\documentclass[a4paper,twoside,12pt]{article}

\usepackage{amsmath}
\usepackage{amssymb}
\usepackage{latexsym}
\usepackage{cite}
\usepackage{graphicx}

\newcommand{\e}{\mathcal{E}}

\newcommand{\g}{\mathcal{G}}

\newcommand{\gthree}{{\gamma_\ast}}

\newcommand{\p}{\mathcal{P}}

\newcommand{\ve}{\varepsilon}

\newcommand{\Ltext}[1]{\ensuremath{\itindex{\mathcal{L}}{#1}}}

\newcommand{\diff}[1][]{\mbox{d}#1}

\newcommand{\half}[1]{\ensuremath{\frac{#1}{2}}}
\newcommand{\intd}[1]{\int \!\! #1 \;}
\newcommand{\inv}[1]{\ensuremath{\frac{1}{#1}}}

\newcommand{\Stext}[1]{\itindex{\mathcal{S}}{#1}}

\newcommand{\itindex}[2]{\ensuremath{#1_{\mbox{\scriptsize{\itshape #2}}}}}

\newcommand{\varfrac}[2][]{\frac{\delta #1}{\delta #2}}

\DeclareMathOperator{\extdm}{d}
\newcommand{\extd}{\extdm \!}

\begin{document}


\renewcommand{\thefootnote}{\fnsymbol{footnote}}
\thispagestyle{empty}
\begin{titlepage}

\begin{flushright}
  TUW-04-29\\
  hep-th/0409283
\end{flushright}
\vspace{1cm}

{\centering \textbf{\large GENERALIZED COMPLEX GEOMETRY AND THE POISSON SIGMA MODEL}\large \par}
\vspace{0.5cm}

\begin{center}
 L. Bergamin\footnote{bergamin@tph.tuwien.ac.at}
\end{center}

{\centering \textit{Institute for Theoretical Physics, Vienna University
of Technology}\par}

{\centering \textit{Wiedner Hauptstra{\ss}e 8-10, A-1040 Vienna, Austria}\par}
\vspace{0.5cm}
\begin{abstract}
The supersymmetric Poisson Sigma model is studied as a possible worldsheet
realization of generalized complex geometry. Generalized complex structures
alone do not guarantee non-manifest $N=(2,1)$ or $N=(2,2)$
supersymmetry, but a certain relation among the different Poisson
structures is needed. Moreover, important relations of an additional
almost complex structure are found, which have no immediate interpretation in terms of generalized
complex structures.
\end{abstract}

\end{titlepage}

%
\renewcommand{\thefootnote}{\arabic{footnote}}
\setcounter{footnote}{0}

\numberwithin{equation}{section}

\section{Introduction}
Two-dimensional non-linear sigma models with extended supersymmetry
\cite{Gates:1983py,Gates:1984nk} recently attracted attention due to the relations to
generalized complex geometry \cite{Hitchin:2003,Gualtieri:2004}. On the one
hand the complex structures of the model of ref.\ \cite{Gates:1984nk} can be
mapped onto two twisted generalized complex structures\cite{Gualtieri:2004,Kapustin:2003sg,Kapustin:2004gv,Zabzine:2004dp,Chiantese:2004pe}. On the other hand it
is expected that generalized complex geometry plays an important role in the first order formulation of the sigma model\footnote{A brief overview of our conventions is given in
  appendix \ref{sec:appendix}. Further details of the notation are explained
  e.g.\ in \cite{Ertl:2000si,Bergamin:2003am}.}
\begin{equation}
  \label{eq:1.1}
  \Ltext{FO} = \half{1} \intd{\diff{^2 \theta}} \bigl( i (D^\alpha X^i) A_{i
  \alpha} - \half{1} \g^{ij}(X) (A_j A_i) - \half{1} \p^{ij}(X) (A_j \gthree A_i)
  \bigr)\ ,
\end{equation}
with metric $\g^{ij}$ and anti-symmetric tensor $\p^{ij}$ as defining
structures of the target space\footnote{A Wess-Zumino type term could be added
to this action as well, however we do not consider such models in the present
work.} \cite{Lindstrom:2004eh,Lindstrom:2004iw}. Here the $X^i$ are scalar
superfields, the $A_{i\alpha}$ real spinorial superfields, i.e.\ their lowest
components $\psi_{i \alpha}$ are Majorana spinors.
The action \eqref{eq:1.1} is manifestly invariant under global $N=(1,1)$
supersymmetry. As the $A_{i\alpha}$ live in $T^*$ it is indeed natural to
expect that additional non-manifest supersymmetries will require a map
$\mathcal{J}$ from $T \oplus T^* \rightarrow T \oplus T^*$ with $\mathcal{J}^2
= - 1$. Such a map is called generalized complex structure, if the natural
indefinite metric on $T \oplus T^*$ is hermitian with respect to $\mathcal{J}$ and
if the latter obeys an integrability
condition with respect to the bracket
\begin{equation}
  [X+\xi,Y+\eta]_C = [X,Y] + \mathcal{L}_X \eta - \mathcal{L}_Y \xi - \half{1}
  \extd(i_X \eta -i_Y \xi)
\end{equation}
for $\mathcal{X}= X+\xi \in T\oplus T^*$
\cite{Hitchin:2003,Gualtieri:2004}. The latter can be written as
\begin{equation}
  [\mathcal{X},\mathcal{Y}]_C -
  [\mathcal{J}\mathcal{X},\mathcal{J}\mathcal{Y}]_C +
  \mathcal{J}[\mathcal{J}\mathcal{X},\mathcal{Y}]_C +
  \mathcal{J}[\mathcal{X},\mathcal{J}\mathcal{Y}]_C = 0
\end{equation}
In the present application it is convenient to write $\mathcal{J}$ in the form
\begin{equation}
  \label{eq:1.3}
  \mathcal{J} = \begin{pmatrix} J & P \\ L & K \end{pmatrix}
\end{equation}
with $J: TM \rightarrow TM$, $P: T^*M \rightarrow TM$, $L: TM \rightarrow
T^*M$ and $K: T^*M \rightarrow T^*M$ ($M$ is now interpreted as the target
space manifold of the action \eqref{eq:1.1}). From $\mathcal{J}^2 = -1$ one
obtains
\begin{align}
\label{eq:1.41}
  J^i{}_j J^j{}_k + P^{ij}L_{jk} &= - \delta^i{}_k\ , \displaybreak[0]\\
\label{eq:1.42}
  J^i{}_j P^{jk} + P^{ij} K_j{}^k &= 0\ , \displaybreak[0]\\
\label{eq:1.43}
  K_i{}^j K_j{}^k + L_{ij} P^{jk} &= - \delta_i{}^k\ , \displaybreak[0]\\
\label{eq:1.44}
  K_i{}^j L_{jk} + L_{ij} J^j{}_k &= 0\ ,
\end{align}
while the hermiticity condition yields
\begin{align}
  \label{eq:1.5}
  J^i{}_j + K_j{}^i &= 0\ , & P^{\{i j\}} &= 0 & L_{\{ij\}} &= 0\ .
\end{align}
Finally the integrability condition can be written as four differential
relations:
\begin{gather}
  \label{eq:1.61}
  J^j{}_{[k}J^i{}_{l],j} + J^i{}_j J^j{}_{[k,l]} + P^{ij} L_{[jk,l]} = 0
  \displaybreak[0]\\
\label{eq:1.62}
  P^{[i|l|} P^{jk]}{}_{,l} = 0\displaybreak[0]\\
  \label{eq:1.63}
  J^i{}_{l,j} P^{jk} + J^i{}_j P^{jk}{}_{,l} + J^k{}_{[l,j]} P^{ij} - J^j{}_l
  P^{ik}{}_{,j} = 0 \displaybreak[0]\\
  \label{eq:1.64}
  J^j{}_i L_{[jk,l]} + J^j{}_{[k} L_{l]i,j} + J^j{}_k L_{jl,i} + L_{ij}
  J^j{}_{[l,k]} + L_{jk} J^j{}_{l,i} = 0
\end{gather}

For the generic
case in \eqref{eq:1.1} the relation between extended supersymmetry and
generalized complex geometry turned out to be very complicated (cf.\
\cite{Lindstrom:2004iw}; notice that these authors start from $(1,0)$
supersymmetry that is different from the model considered here). In
ref.\ \cite{Lindstrom:2004eh} the problem has been solved by going partly
on-shell together with the restriction that $\g$ must be invertible. Both
works do not cover the Poisson Sigma
model (PSM) \cite{Schaller:1994es,Schaller:1994uj,Cattaneo:1999fm}, which has important applications in string theory due to
the path integral interpretation \cite{Cattaneo:1999fm} of Kontsevitch's
$\star$-product \cite{Kontsevich:1997vb}. In addition, many questions remained
open in \cite{Lindstrom:2004eh,Lindstrom:2004iw}
and a simple example as the PSM certainly can help to clarify the situation.
Thus we consider in this work the action \eqref{eq:1.1} in the limit of a PSM, where $\g
= 0$ and $\p$ is Poisson. In
particular, the relation between PSMs with generalized complex target space
and PSMs with non-manifest extended supersymmetry is analyzed. It should be
noted that supersymmetric extensions of the PSM with \emph{manifest} $N=(4,4)$
supersymmetry have been obtained in \cite{Ivanov:1995jb,Ivanov:1995yp} using
harmonic superspace techniques.

To perform the calculations it is preferable to write the action
\eqref{eq:1.1} in chiral components. With $\e = \g + \p$ this becomes
\begin{equation}
  \label{eq:1.2}
  \Ltext{FO} = \half{1} \intd{\diff{^2 \theta}} \bigl( i (D_- X^i) A_{i
  +} - i (D_+ X^i) A_{i -} + \e^{ij} A_{j +} A_{i -}
  \bigr)\ .
\end{equation}
The strategy to find non-manifest extended supersymmetry is now quite easy:
One writes down the most general transformations for the component fields in
\eqref{eq:1.2}. Then the invariance of the action under these
transformations is checked, which will lead to a number of algebraic and differential relations. Finally
the new supersymmetries must obey the supersymmetry algebra
\eqref{eq:1.11} and moreover commute with the manifest supersymmetry. The last point
is automatically satisfied if the transformations are written in a superspace
covariant form.

For an additional supersymmetry with transformation parameter
$\ve^+$, yielding a $N=(2,1)$ invariant theory, dimensional analysis tells us that
\begin{gather}
  \label{eq:1.71}
  \delta_+ X^i = \ve^+ D_+ X^j J^{+i}{}_j + \ve^+ A_{j+} P^{+ij}\ , \displaybreak[0]\\
  \label{eq:1.72}
\begin{split}
  \delta_+ A_{i+} &= \sqrt{2} \ve^+ \partial_{++} X^j L^+_{ij} - \ve^+ D_+ A_{j+}
  K^+_i{}^j + \ve^+ D_+ X^j D_+ X^k M^+_{ijk}\\&\quad  + \ve^+ A_{j+} A_{k+} N^+_i{}^{jk} +
  \ve^+ D_+ X^j A_{k+} Q^+_{ij}{}^k\ ,
\end{split} \displaybreak[0]\\
  \label{eq:1.73}
\begin{split}
  \delta_+ A_{i-} &= \ve^+ D_+ A_{j-} R^+_i{}^j + \ve^+ D_- A_{j+} S^+_i{}^j +
  \ve^+ A_{j+} A_{k-} Y^+_i{}^{jk} + \ve^+ A_{j +} D_- X^k U^+_{ik}{}^j  \\
  &\quad   + \ve^+ D_+ X^j A_{k-} V^+_{ij}{}^k +
  \ve^+ D_+ D_- X^j T^+_{ij} + \ve^+ D_+ X^j D_- X^k W^+_{ijk}\ .
\end{split}
\end{gather}
Analogous relations follow for an additional supersymmetry $Q_-$.

Before starting the calculations the relation to the more familiar second
order formulation of the action \eqref{eq:1.1} or \eqref{eq:1.2} is outlined
briefly. If $\e^{ij}$ is invertible the spinorial fields $A_{i \alpha}$ can be eliminated by the
equations of motion from \eqref{eq:1.2}
\begin{align}
\label{eq:NN}
  \varfrac{A_{i+}} \Stext{FO} &= - \half{i} D_- X^i + \half{1} \e^{ji} A_{j-}\ ,
  & \varfrac{A_{i-}} \Stext{FO} &=  \half{i} D_+ X^i - \half{1} \e^{ij}
  A_{j+}\ .
\end{align}
Then the second order Lagrangian ($\e_{ij} = (\e^{-1})_{ij}$)
\begin{equation}
  \label{eq:2.24}
  \Ltext{SO} = \half{1} \intd{\diff{^2 \theta}} D_+ X^i D_- X^j \e_{ji} =
  \half{1} \intd{\diff{^2 \theta}} D_+ X^i D_- X^j (g + b)_{ji}
\end{equation}
with the corresponding symmetry transformations
\begin{align}
\label{eq:2.251}
  \delta_+ X^I &= \ve^+ D_+ X^j I^{+ i}{}_j & I^+ &= J^+ + i P^+ \e^{-1}  \displaybreak[0]\\
\label{eq:2.252}
  \delta_- X^I &= \ve^- D_- X^j I^{- i}{}_j & I^- &= J^- - i (\e^{-1} P^-)^T 
\end{align}
is obtained.
These transformations define two additional non-manifest supersymmetries, if
the metric $g$ is hermitian with respect to $I^{\pm}$ and if $I^{\pm}$ define
covariantly constant
complex structures \cite{Gates:1984nk}.
The definition of $(g, H = \extd b, I^+, I^-)$ is equivalent to a certain
$H$-twisted generalized K\"{a}hler structure on $M$ \cite{Gualtieri:2004}. 
\section{General Sigma Model}
In a first step the invariance of the action and one relation of the
supersymmetry algebra are analyzed. We exemplify the calculations at hand of
the $\delta_+$ transformation, the generalization to $\delta_-$ is
straightforward. For simplicity, the superscript ``+'' for the various tensors
in \eqref{eq:1.71}-\eqref{eq:1.73} is omitted in this section. Moreover, we do not yet impose any constraints on
$\e^{ij}$. Most results of this section can be found in \cite{Lindstrom:2004iw}
already, they are reproduced here to clarify our notations and
conventions. However, ref.\ \cite{Lindstrom:2004iw} in many formulas assumes
an invertible $\e$, a constraint that we do not impose here.

The invariance of the action \eqref{eq:1.2} under
\eqref{eq:1.71}-\eqref{eq:1.73} can now be studied order in order
of the gauge potentials $A_{i \alpha}$. Going through all steps one finds eleven different conditions:
\begin{gather}
  \label{eq:2.2}
  L_{\{ij\}} = -  T_{\{ij\}} \displaybreak[0]\\
  \label{eq:2.3}
  M_{ijk} - W_{jki} = \inv{2}
  L_{[ij,k]} - \half{1} T_{jk,i}\displaybreak[0]\\
\label{eq:2.4}
  J^j{}_i + K_i{}^j = S_i{}^j - i \e^{kj} T_{ki} \displaybreak[0]\\
  \label{eq:2.5}
  K_{[k}{}^i{}_{,j]} + Q_{kj}{}^i - U_{jk}{}^i = i \bigl((\e^{li} T_{lj})_{,k} -
  \e^{li} W_{ljk} \bigr)\displaybreak[0]\\
  \label{eq:2.6}
  J^j{}_i + R_i{}^j = -i \e^{jk} L_{ki} \displaybreak[0]\\
  \label{eq:2.7}
  V_{[ij]}{}^k = i \bigl( \e^{kl}M_{l[ij]}- (\e^{kl} L_{l[i})_{,j]}\bigr) \displaybreak[0]\\
  \label{eq:2.8}
  \e^{ij} N_j{}^{kl} = \half{1} \bigl(\e^{j[k} Y_j{}^{l]i} - \e^{i[l}{}_{,j} P^{j|k]}\bigr)\displaybreak[0]\\
  \label{eq:2.9}
  P^{\{ij\}} = i \e^{k\{i} S_k{}^{j\}}\displaybreak[0]\\
  \label{eq:2.10}
  P^{ij} = -i (\e^{ik} K_k{}^j + \e^{kj} R_k{}^i) \displaybreak[0]\\
  \label{eq:2.11}
  2 N_i{}^{jk} =  i \e^{l[j} U_{li}^{k]} - \half{1} P^{[kj]}{}_{,i} + \half{i}
  (\e^{l[j} S_l{}^{k]})_{,i} \displaybreak[0]\\
  \label{eq:2.12}
  Y_i{}^{jk} = i \bigr( \e^{lj} V_{li}{}^k + (\e^{kl} K_l{}^j)_{,i} +
  \e^{kl} Q_{li}{}^j + \e^{kj}{}_l J^l{}_i) \bigl)
\end{gather}
In particular we obtain in the limit of $\e \rightarrow 0$
\begin{align}
  \label{eq:2.13}
  R_i{}^j &= - J^j{}_i\ , & P^{ij}&= V_{[ij]}{}^k = N_i{}^{jk} = Y_i{}^{jk} = 0\ .
\end{align}
More involved is the calculation of the supersymmetry algebra. The definition $\delta_+ \Psi = i [\ve^+ Q_+, \Psi]$
implies that
\begin{equation}
  \label{eq:2.14}
  \delta_+^1 \delta_+^2 \Psi = \sqrt{2} \ve_1^+ \ve_2^+ \partial_{++} \Psi\ .
\end{equation}

We start with the commutator acting on $X^i$.
First of all the fact that the derivative term on the rhs of \eqref{eq:2.14}
is generated yields as condition exactly eq.\ \eqref{eq:1.41}. Therefore,
three out of four components of the generalized complex structure are already
identified. Consequently, we also \emph{choose} $L_{\{ij\}} = P^{\{ij\}} = 0$,
else we do not obtain the desired structure\footnote{The symmetric part of $L$
and $T$ can be set to zero by means of a ``field equation'' symmetry and an
appropriate redefinition of various quantities in \eqref{eq:1.72} and
\eqref{eq:1.73}. In general, this is
not possible for the symmetric part of $P$, as the corresponding term $\propto P A_+$ in
\eqref{eq:1.71} is not part of a field equation.}. Next we consider terms $\propto D_+ X^i D_+
X^j$. If
\begin{equation}
  \label{eq:2.15}
  M_{ijk} = \inv{2} L_{[ij,k]}
\end{equation}
these terms are equivalent to \eqref{eq:1.61}. The terms $\propto D_+ A_{i+}$
lead to \eqref{eq:1.42} and this identifies the last map of $\mathcal{J}$. The
remaining conditions can be identified with \eqref{eq:1.62} and
\eqref{eq:1.63} if
\begin{align}
\label{eq:2.16}
  Q_{ij}{}^k &= J^k{}_{[i,j]}\ , & N_i{}^{kl} &= \half{1} P^{kl}{}_{,i}\ .
\end{align}
Thus the condition that $\delta_+^1 \delta_+^2 X^i = \sqrt{2} \ve_1^+
\ve_2^+ \partial_{++} X^i$, under the assumption that the target-space is
generalized complex, constrains the transformation of $A_{i+}$ as
\begin{equation}
  \label{eq:2.17}
\begin{split}
  \delta_+ A_{i +} &= \sqrt{2} \ve^+ \partial_{++} X^j L_{ij} + \ve^+ D_+ A_{j+}
  J^j{}_i + \half{1} \ve^+ D_+  X^j D_+ X^k L_{[ij,k]}\\&\quad  + \half{1} \ve^+ A_{j+} A_{k+} P^{jk}{}_{,i} +
  \ve^+ D_+ X^j A_{k+} J^k{}_{[i,j]}\ .
\end{split}
\end{equation}
Up to conventions this is exactly the result of \cite{Lindstrom:2004iw}. As
$\delta_+^1 \delta_+^2 A_{i+}$ is obviously independent of $A_{i-}$ we can borrow
the result therefrom that this transformation satisfies the supersymmetry
algebra relation if the target space is generalized complex.

Now we should go again through the conditions
\eqref{eq:2.2}-\eqref{eq:2.12}. From \eqref{eq:2.2}-\eqref{eq:2.5},
\eqref{eq:2.7} and \eqref{eq:2.12} follows
\begin{gather}
  \label{eq:2.18}
\begin{alignat}{3}
  T_{\{ij\}} &= 0\ , &\qquad W_{ijk} &= \half{1} T_{ij,k}\ , &\qquad S_i{}^j &= i \e^{kj}
  T_{ki}\ ,
\end{alignat}
  \displaybreak[0]\\
  \label{eq:2.19}
\begin{alignat}{2}
  U_{ij}{}^k &= - \half{i} \e^{lk} T_{li,j} - i \e^{lk}{}_{,j} T_{li} \ , &\qquad V_{[ij]}{}^k &= i \e^{kl}
  L_{ij,l}- i\e^{kl}{}_{,[i} L_{j]l} \ ,
\end{alignat} \displaybreak[0]\\
\label{eq:2.19.2}
 Y_i{}^{jk} = i(\e^{lj} V_{li}{}^k - \e^{kl} J^j{}_{i,l} - \e^{kl}{}_{,i}
 J^j{}_{l} + \e^{kj}{}_{,l} J^l{}_i)\ .
\end{gather}
Furthermore, the relations \eqref{eq:2.9} and \eqref{eq:2.11} are
automatically satisfied.

At this point we should have a careful look at the definition of $P$ in
\eqref{eq:2.10}. Together with \eqref{eq:2.6} and the split $\e = \g + \p$ one
finds after some algebra that
\begin{equation}
  \label{eq:2.20}
  P^{ij} = -i (\p K)^{[ij]} - (\p L \p)^{ij} - (\g L \g)^{ij}
\end{equation}
is the correct definition of the anti-symmetric tensor $P$. However, the definition
\eqref{eq:2.10} contains a symmetric part as well, which must vanish. This
leads to the constraint
\begin{equation}
  \label{eq:2.21}
  i (\g K)^{\{ij\}} + (\p L \g)^{ij} + (\g L \p)^{ij} = 0\ .
\end{equation}
It is seen that the constraint is trivial for $\g = 0$, on the other hand
it reduces for $\p = 0$ to the condition that the metric $\g$ is hermitian with respect to
$K$.

As a side remark we mention that under the elimination of $A_{i\alpha}$ in eq.\
\eqref{eq:NN} $I^+ =
  \e(K^+ \e^{-1} - i L^+)$ automatically squares to $-1$ due to the
  constraints derived in this section (cf.\ also \cite{Lindstrom:2004iw}). If $\p=0$ it is
  easily seen that $[\mathcal{J}^+,\mathcal{J}^-] = 0$ implies $[I^+,I^-] =
  0$, i.e.\ in this special case there exists a
simple relation between the generalized complex
structures of the first order model and of the generalized K\"{a}hler
structure resp.

\section{Supersymmetric PSM}
In the current work we want to investigate the case where the action
\eqref{eq:1.1} reduces to a supersymmetric PSM. This allows an important
simplification of the analysis: as the Poisson tensor can be transformed
to Casimir-Darboux coordinates it is sufficient to consider a constant $\p$ for any analysis local in the
target space manifold. Thus we consider in the following the case where $\e$
reduces to a constant and anti-symmetric matrix, denoted by $\p$.

As is obvious from \eqref{eq:1.5} and \eqref{eq:1.63} $P$ is a Poisson
tensor. It is crucial to distinguish this tensor clearly from $\p$ in
\eqref{eq:1.1}. In the current situation, $\p$ is Poisson as well, but the two
Poisson structures are not equivalent, though they are
related. In particular
it follows, that the rank of $P$ cannot be larger than the rank of $\p$, but
$P$ need not be constant for constant $\p$. Thus the relations
\eqref{eq:1.41}, \eqref{eq:1.42} and \eqref{eq:1.61}-\eqref{eq:1.63} should be
investigated with the restriction $P^{ij} =  -i (\p K)^{[ij]} - (\p L
\p)^{ij}$. Eq.\ \eqref{eq:1.41} is most elegantly written as
\begin{equation}
  \label{eq:2.40}
  (R^2)_i{}^j = - \delta_i{}^j\ ,
\end{equation}
i.e.\ $R$ is an almost complex structure. Then with the use of \eqref{eq:1.44}
eq.\ \eqref{eq:1.42} is satisfied identically. Eqs.\
\eqref{eq:1.61}-\eqref{eq:1.63} yield complicated differential conditions that
we do not reproduce in full generality here. Notice that the relation
\eqref{eq:2.8}, the only restriction from the invariance of the action that we
did not solve in the previous section, is satisfied identically in the case of
the PSM.

Now, the derivation of the remaining commutators splits into to parts: First we
consider the extension to $N=(2,1)$ supersymmetry. There the only remaining
commutator is $[\delta^1_+, \delta^2_+] A_{i-}$. Then one has to ensure that
the commutators $[\delta_+, \delta_-] \Psi$ vanish for all fields $\Psi$.
\subsection{$\mathbf{N=(1,1) \rightarrow N=(2,1)}$}
While the the commutators $[\delta^1_+, \delta^2_+] \Psi$ could be solved for
$X^i$ and $A_{i+}$ off-shell in full generality, unraveling the structure of a
generalized complex geometry at the target space, this does not seem to be
possible for the remaining field $A_{i-}$ (cf.\ the complicated relations in
\cite{Lindstrom:2004iw}, esp.\ eq.\ (A.3)). Indeed, it is not expected that
the non-manifest supersymmetry closes off-shell, except for certain very
special cases. Therefore, only on-shell closure of the algebra will be
demanded in the following and consequently the
transformations \eqref{eq:1.71}-\eqref{eq:1.73} together with the restrictions
derived in the previous section can be reduced to
\begin{align}
  \label{eq:N2}
  \delta_+ X^i &= - i \ve^+ A_{j+} (\p R)^{ij}\ , \displaybreak[0]\\
  \label{eq:N3}
  \delta_+ A_{i+} &= - \ve^+ D_+ A_{j+} R_i{}^j - i \ve_+ A_{j+} A_{k+}
  \p^{jl} R_i{}^k{}_{,l}\ , \displaybreak[0]\\
  \label{eq:N4}
  \delta_+ A_{i-} &=  \ve^+ D_+ A_{j-} R_i{}^j - i \ve_+ A_{j-} A_{k+}
  \p^{jl} R_i{}^k{}_{,l}\ .
\end{align}
Not surprisingly, the symmetry transformations now can be written in terms of
a single almost complex structure, which is in agreement with previous results
\cite{Gates:1984nk,Lindstrom:2004eh}. Nevertheless, notice the difference to
these approaches. As we do not insist on $\p$ having full rank, the equations
of motion cannot be solved for $A_{i\alpha}$ (cf.\ \eqref{eq:NN}) and the transformation
rule for $X^i$ cannot be reduced to the form \eqref{eq:2.251}. Of course the
relation \eqref{eq:N2} looks rather strange, in
particular one obtains $\delta X^i = 0$ for BF theory. But in that case $D_\alpha X^i$
already is an equation of motion and therefore the representation of
supersymmetry on that field vanishes on-shell.

Now, the derivation of $[\delta^1_+, \delta^2_+] A_{i-}$ is surprisingly
easy. As $R$ is almost complex, the correct supersymmetry algebra is generated
in an obvious way. All remaining contribution are found to vanish if the modified integrability
condition
\begin{equation}
  \label{eq:N5}
  \p^{m[k} R_i{}^{|j|} R_j{}^{l]}{}_{,m} - \p^{mj} R_j{}^{[k} R_i{}^{l]}{}_{,m} = 0
\end{equation}
holds. First notice that this relation is satisfied for any Poisson tensor
$\p$ if $R \equiv K$, i.e.\ $L \equiv 0$. A simple interpretation can be given
if $\p$ can be used as an intertwiner that defines a new map $TM
\rightarrow TM$ as\footnote{A similar observation has been made in
  \cite{Lindstrom:2004iw} as well, but these authors used a different intertwiner.}
\begin{equation}
  \label{eq:NN1}
  \p R = \tilde{I} \p\ .
\end{equation}
In the case of a symplectic $\p$ the new almost complex structure $\tilde{I}$
is exactly $I$ of eq.\ \eqref{eq:2.251}. Then the differential condition
\eqref{eq:N5} is nothing but the integrability condition (vanishing Nijenhuis
tensor) of $I$ already found in \cite{Gates:1984nk}. If $\p$ does not have
full rank, $I$ does no longer exist, but on each symplectic leaf a similar
structure still can be defined. By choosing $\tilde{I}$ to be this almost
complex strucutre, \eqref{eq:N5} reduces the integrability condition for $\tilde{I}$
if $R$ is of the form $R = R_0 \oplus R_1$, where $R_0$ lives in the
symplectic leaf, while $\p R_1 \equiv 0$. Furthermore,
it can be checked that \eqref{eq:N5} is identically zero for all components of
$R_1$. It should be stressed that this is not the most general solution of
\eqref{eq:N5} as $R$ needs not be decomposable in this way.

The tensors $T_{ij}$ and $V_{ij}{}^k$ remain
undetermined. As all components of $\delta_+ A_{i+}$ and $\delta_+ X^i$ have
been fixed by the requirement that the target-space is generalized complex
this ambiguity is not related to ``field equation'' symmetries. Of course,
the off-shell closure of some transformations is somehow arbitrary, if this
constraint is not imposed for all fields. However, this is sufficient to analyze
under which conditions the sigma model, whose target space is equipped with a
generalized complex structure, has an additional non-manifest supersymmetry.

For completeness the conditions for off-shell closure are reproduced for the
special case of BF theory. In this simple model $P\equiv0$, $R \equiv K$ and the 
correct supersymmetry algebra is ensured by the fact that $J^2 = -1$. All
other contributions have to vanish. The relations involving $T$ can be cast
into the form
\begin{gather}
  \label{eq:3.24}
   T_{kj}J^j{}_i  + T_{ij}J^j{}_k  = 0\ , \displaybreak[0]\\
\label{eq:3.25}
\begin{alignat}{2}
   T_{ij} J^j{}_{k,l} -  T_{kj} J^j{}_{i,l} &= 0\ , &\qquad
  (T_{kj} J^j{}_{[i})_{,l]} &=  T_{[i|k,j} J^j{}_{l]}\ , 
\end{alignat}\displaybreak[0]\\
\label{eq:3.26}
V_{\{il\}}{}^j T_{jk} = J^j{}_{\{i} T_{l\}k,j} + (T_{jk} J^j{}_{\{i})_{,l\}} -
\half{1} J^j{}_{\{i|,k} T_{j|l\}}\ , \displaybreak[0]\\
\label{eq:3.27}
  (J^j{}_{[k,l]} T_{ji} )_{,m} + (J^j{}_{[k} T_{l]i,j})_{,m} + J^{j}{}_i
  T_{j[k,l]m} + J^j{}_{[k,l]m} T_{ji} =  V_{i[k}{}^j T_{l]j,m}\ . 
\end{gather}
Obviously, $T\equiv 0$ is a simple and appealing solution for this case. The
other immediate guess $T \propto L$ is not possible in general, as the above
set of differential conditions does not reduce to
\eqref{eq:1.41}-\eqref{eq:1.64}.
 
The remaining conditions yield algebraic and differential equations for $V$:
\begin{gather}
  \label{eq:3.28}
  J^j{}_{[i} V_{k]j}{}^l = 0 \displaybreak[0]\\
  \label{eq:3.29}
  J^k{}_{j,l} J^j{}_i + J^k{}_{i,j} J^j{}_l  = J^k{}_j V_{il}{}^j -  J^j{}_i
  V_{jl}{}^k\displaybreak[0]\\
  \label{eq:3.30}
  -J^j{}_i V_{j[k}{}^m{}_{,l]} + J^j{}_{[k,l]} V_{ij}{}^m + J^j{}_{[k}
   V_{l]i}{}^m{}_{,j} = V_{i[k}{}^j V_{l]j}{}^m
\end{gather}
Notice the similarity between \eqref{eq:3.28} and the condition
\eqref{eq:1.61} for $P=0$: Indeed, with \eqref{eq:2.13} and \eqref{eq:2.16} we
can write the latter as $J^j{}_{[i} Q_{k]j}^{}l = 0$. However, $Q_{ij}{}^k$
is antisymmetric in its lower indices while $V_{ij}{}^k$ is symmetric. Due to
this characteristic, the rhs of \eqref{eq:3.29} is found to vanish when
anti-symmetrized in $i$ and $l$. But this does not lead to a new constraint
for $J$, as the lhs is found to reduce to \eqref{eq:1.61} in that case.

One
might be tended to choose $V_{ij}{}^k = 0$ as well. But this yields an
additional constraint onto the generalized complex structure, as the
integrability condition \eqref{eq:1.61} must split into two independent pieces
according to \eqref{eq:3.29}.

We do not go into further details about the off-shell closure of the
PSM. Notice however, that $T\equiv0$ will again be a solution of the $N=(2,1)$
extension. On the other hand, the differential conditions for $V$ become much more involved
than in the case above. Unfortunately, it does not seem to be possible to
bring them into a form similar to \eqref{eq:3.28}-\eqref{eq:3.30} by substituting
$J\rightarrow -R^T$, as one might expect naively.
\subsection{$\mathbf{N=(1,1) \rightarrow N=(2,2)}$}
To implement $N=(2,2)$ supersymmetry the result found so far for $\delta_+$ must
be generalized to $\delta_-$. But this is almost trivial: first one has to
exchange all indices $+ \leftrightarrow -$, moreover the indices of all
tensors $\e$ have to be exchanged: $\e^{ij} \rightarrow \e^{ji}$. For the
special case of the PSM this boils down to add a sign in front of every
$\p^{ij}$.

Now the tensors from $\delta_+$ and
$\delta_-$ must be distinguished again by the use of the labels according to
\eqref{eq:1.71}-\eqref{eq:1.73}. Then the two supersymmetries
are found to commute if the two conditions
\begin{gather}
\label{eq:4.1}
  [R^+, R^-] = 0\ , \displaybreak[0]\\
\label{eq:4.2}
  \p^{km} R^+_i{}^j R^-_j{}^l{}_{,m} - \p^{lm} R^-_i{}^j R^+_j{}^k{}_{,m} +
  (\p R^+)^{mk} R^-_i{}^l{}_{,m} - (\p R^-)^{ml} R^+_i{}^k{}_{,m} = 0
\end{gather}
hold. Notice that \eqref{eq:4.2} reduces to \eqref{eq:N5} for $R^+ =
R^-$. Furthermore, for symplectic $\p$ these conditions should reduce to the ones found in
\cite{Gates:1984nk} (cf.\ also \cite{Lindstrom:2004eh}). It is interesting to
study the relation of \eqref{eq:4.1} to a possible constraint
$[\mathcal{J}^+,\mathcal{J}^-] = 0$. With the definition of $R$ in eq.\
\eqref{eq:2.6} the commutator \eqref{eq:4.1} becomes (recall the different
sign in the definition of $R^-$)
\begin{equation}
  \label{eq:4.3}
  [K^+,K^-] + [L^+\p,L^-\p] + i [K^+, L^-\p] - i [L^+\p, K^-] = 0\ ,
\end{equation}
while the relevant commutator from $[\mathcal{J}^+,\mathcal{J}^-] = 0$ is
found to be
\begin{equation}
  \label{eq:4.4}
  [K^+,K^-] - [L^+\p,L^-\p] + i (L^+ \p K^- - L^- J^- \p + L^- \p K^+ - L^-
  J^+ \p) = 0\ .
\end{equation}
The difference of these two equations does not vanish by means of the
remaining conditions form $[\mathcal{J}^+,\mathcal{J}^-] = 0$.
Therefore we conclude that the constraints \eqref{eq:4.1} and \eqref{eq:4.2}
cannot be interpreted in straightforward way as parts of the generalized
complex structure.

It may be useful to summarize at this point the conditions derived for an
$N=(2,2)$ supersymmetric PSM: It was assumed that $J^\pm$, $P^\pm$, $L^\pm$ and $K^\pm$ in
\eqref{eq:1.71} and \eqref{eq:1.72} define two generalized complex
structures. Then the transformations of $A_{i\pm}$ under $\delta_\pm$ are given
by \eqref{eq:2.17}. The remaining transformations $\delta_\pm A_{i\mp}$ have
been studied for on-shell closure, only. These transformations depend on two
almost complex structures $R^\pm$ that must satisfy \eqref{eq:N5},
\eqref{eq:4.1} and \eqref{eq:4.2}.
\section{Conclusions}
In this work the relation between extended non-manifest supersymmetry of the
Poisson Sigma model and generalized complex structures has been studied. As
expected, not every PSM with extended supersymmetry constrains its target
space to be generalized complex. The converse is true neither: beside the
relations among the Poisson structures of generalized complex geometry and of
the PSM resp., it was found that an additional almost complex structure must
obey a (modified) integrability condition, which in certain cases can be interpreted as a
complex structure restricted to the symplectic leaves of the Poisson manifold. Finally the
conditions for $N=(2,2)$ supersymmetry have been analyzed, the ensuing
conditions do not necessarily imply that the two generalized complex
structures commute.

Among the unanswered question there remains the interpretation of certain
differential conditions, esp.\ the constraint of vanishing Nijenhuis tensor of
the Poisson structure $P$ in \eqref{eq:1.62}. Also, the conditions for
off-shell closure of the algebra could not be solved. Here important
additional constraints on the target space appear that we were not yet able to
interpret in a conclusive way.

Of course, it would be interesting to study extensions of the
model, e.g.\ the inclusion of a Wess-Zumino term \cite{Klimcik:2001vg}, where
a twisted generalized complex structure is expected \cite{Lindstrom:2004iw}, or
non-topological extensions. Finally, we did not consider global effects, such
as changes of the rank of the Poisson tensor $\p$ or effects from
non-trivial boundary conditions.

\section*{Acknowledgement}

It is a pleasure to thank E.~Scheidegger, H.~Balasin and T.~Strobl for numerous
important discussions on (generalized) complex geometry. Also I would like to
thank W.~Kummer for helpful comments.
This work has been supported by the project P-16030-N08 of the
Austrian Science Foundation (FWF).

\appendix
\section{Notations and Conventions}
\label{sec:appendix}
The conventions used are explained in detail in
\cite{Ertl:2000si,Bergamin:2003am}. As they differ from the ones in
\cite{Lindstrom:2004eh,Lindstrom:2004iw} the most important definitions are
summarized in this appendix.

The $\gamma$-matrices are used in a chiral representation:
\begin{align}
\label{eq:gammadef}
  {{\gamma^0}_\alpha}^\beta &= \left( \begin{array}{cc} 0 & 1 \\ 1 & 0
  \end{array} \right) & {{\gamma^1}_\alpha}^\beta &= \left( \begin{array}{cc} 0 & 1 \\ -1 & 0
  \end{array} \right) & {{\gthree}_\alpha}^\beta &= {(\gamma^1
    \gamma^0)_\alpha}^\beta = \left( \begin{array}{cc} 1 & 0 \\ 0 & -1
  \end{array} \right)
\end{align}
 As we work with spinors in a
chiral representation we can use $ \chi^\alpha = ( \chi^+, \chi^-)$
and upper and lower chiral components are related by $\chi^+
= \chi_-$, $ \chi^- = - \chi_+$, $\chi^2 = \chi^\alpha \chi_\alpha = 2 \chi_- \chi_+$.
Furthermore for Majorana spinors $\chi^+$ is real while $\chi^-$ is
imaginary. Vectors in light-cone coordinates coincide with the respective spin
tensor decomposition if they are defined as $v^{++} = \frac{i}{\sqrt{2}} (v^0
+ v^1)$, $v^{--} = \frac{-i}{\sqrt{2}} (v^0 - v^1)$.

Finally the basic conventions of $(1,1)$ supersymmetry are explained. The
representation of the supercharges is chosen as
\begin{align}
\label{eq:1.11}
  Q_\alpha &= \partial_\alpha - i (\gamma^a \theta)_\alpha \partial_a\ , & \{
  Q_\alpha, Q_\beta \} &= 2 i \gamma^a_{\alpha \beta} \partial_a\ ,
\end{align}
which yields as a convenient choice of the supersymmetry-covariant derivatives
$D_\alpha = \partial_\alpha + i (\gamma^a \theta)_\alpha \partial_a$.
In chiral components these derivatives obey
\begin{align}
  \label{eq:app1}
  \{ D_+, D_- \} &= 0\ , & D^2_+ &= - \sqrt{2} \partial_{++}\ , & D^2_- &= -
  \sqrt{2} \partial_{--}\ .
\end{align}

\providecommand{\href}[2]{#2}\begingroup\raggedright\endgroup

\end{document}